\documentclass[letter]{aa} % for the letters 
\usepackage[varg]{txfonts}
\usepackage{graphicx}%,subfigure}
\usepackage{epsfig}
\usepackage{color}
\usepackage{amssymb}
\usepackage{txfonts}
\usepackage{lscape}
\usepackage{float}
\usepackage{amsmath}
\usepackage{color}
\usepackage{longtable}
\usepackage{rotating}
\usepackage{pdflscape}
\usepackage{dcolumn}
\usepackage{caption}
\usepackage{subcaption}
\usepackage[nameinlink]{cleveref}
\usepackage[normalem]{ulem}

\usepackage{color,soul}
\usepackage{amssymb}
\usepackage{lscape}
\usepackage{float}
\usepackage{enumitem}
\usepackage{mathtools}
\DeclareMathOperator{\MyProd}{\scalebox{1.4}{$\mathrm{I\kern-0.2ex I}$}}
\usepackage{subcaption}
\usepackage{caption}
\usepackage{xcolor}
\usepackage[export]{adjustbox}
\DeclareCaptionLabelFormat{continued}{Figure \ref{postage}}
\def\la{\mathrel{\hbox{\rlap{\hbox{\lower4pt\hbox{$\sim$}}}\hbox{$<$}}}}
\def\ga{\mathrel{\hbox{\rlap{\hbox{\lower4pt\hbox{$\sim$}}}\hbox{$>$}}}}

\def\arcmin{\hbox{$^\prime$}}
\def\arcsec{\hbox{$^{\prime\prime}$}}

\newcommand{\dg}{^{\circ}}

\newcommand{\MSUN}{${\rm M}_\odot$}

\newcommand{\kms}{{\,km\,s$^{-1}$}}

\def\apjl   {{ApJL }}
\def\apjs   {{ApJS }}

\begin{document}

\title{Collimated synchrotron threads linking the radio lobes of ESO~137-006}

\subtitle{}
\author{M. Ramatsoku\inst{1,}\inst{2}\fnmsep\thanks{m.ramatsoku@ru.ac.za} \and M. Murgia\inst{2} \and V. Vacca\inst{2} \and P. Serra\inst{2} \and S. Makhathini\inst{1} \and F. Govoni\inst{2} \and O. Smirnov \inst{1,}\inst{3} \and L. A. L. Andati\inst{1} \and E. de Blok\inst{7,5,6} \and G. I. G. J\'{o}zsa\inst{3,1,4} \and P. Kamphuis\inst{9} \and D. Kleiner\inst{2} \and F. M. Maccagni\inst{2} \and D. Cs. Moln\'{a}r\inst{2} \and A. J. T. Ramaila\inst{3} \and K. Thorat\inst{8} \and S.V. White\inst{1}.}

\institute{Department of Physics and Electronics, Rhodes University, PO Box 94, Makhanda, 6140, South Africa. 
         \and
             INAF- Osservatorio Astronomico di Cagliari, Via della Scienza 5, I-09047 Selargius (CA), Italy.
         \and 
           South African Radio Astronomy Observatory, 2 Fir Street, Black River Park, Observatory, 7925, South Africa.             
         \and
           Argelander-Institut fur Astronomie, Auf dem Hugel 71, D-53121 Bonn, Germany.
          \and 
           ASTRON, Netherlands Institute for Radio Astronomy, Oude Hoogeveensedijk 4, 7991 PD, Dwingeloo, The Netherlands.
          \and 
           Department of Astronomy, University of Cape Town, Private Bag X3, Rondebosch 7701, South Africa.
          \and
           Kapteyn Astronomical Institute, University of Groningen, Postbus 800, 9700 AV, Groningen, The Netherlands.
          \and 
           Department of Physics, University of Pretoria, Hatfield, Pretoria, 0028, South Africa.
          \and 
           Ruhr University Bochum, Faculty of Physics and Astronomy, Astronomical Institute, 44780 Bochum, Germany. 
           }

 \date{Received 24 February 2020; Accepted 09 March 2020}

% \abstract{}{}{}{}{} 
% 5 {} token are mandatory
 
  \abstract{We present MeerKAT 1000\,MHz and 1400\,MHz observations of a bright radio galaxy in the southern hemisphere, ESO~137-006. The galaxy lies at the centre of the massive and merging Norma galaxy cluster. The MeerKAT continuum images (rms $\sim$0.02 mJy/beam at $\sim$10\arcsec resolution) reveal new features that have never been seen in a radio galaxy before: collimated synchrotron threads of yet unknown origin, which link the extended and bent radio lobes of ESO~137-006. The most prominent of these threads stretches in projection for about 80 kpc and is about 1 kpc in width. The radio spectrum of the  threads is steep, with a spectral index of up to $\alpha\simeq 2$ between 1000 and 1400\,MHz. 
  }


\maketitle

%
%________________________________________________________________

\section{Introduction}
\label{sec:intro}
Radio galaxies residing in galaxy clusters exhibit a wide range of distorted morphologies (see \citealp{Garon2019} for an overview). Some of these distortions are seen in the form of head tails, which are elongated radio sources with the galaxy at one end (\citealp{Ryle1968}, \citealp{Sebastian2017}), and wide-angle tails, where the jets form a `C' shape (\citealp{Owen1976}, \citealp{Leahy1993}, \citealp{Missaglia2019}). Such radio morphologies are due to the interaction between the radio lobes and/or jets and the intra-cluster medium (ICM; \citealp{Pinkney1995}, \citealp{Sakelliou2000}). This makes radio galaxies important probes of the distribution of pressure, turbulence, and shocks within the magneto-ionic ICM (\citealp{Owen2014}, \citealp{Feretti2012}), and also means that their detection is an efficient way to identify high-z clusters independent of dust extinction.

In this letter we discuss the case of ESO~137-006, a radio galaxy in the Norma galaxy cluster (Abell~3627; \citealp{Abell1989}). Norma is located at a distance of $\sim70$ Mpc ($z$ = 0.0162; \citealp{Woudt2008}) at the crossing between several filaments in the Great Attractor region \citep{Dressler1987GA}. With a dynamical mass of $\sim 10^{15}$ \MSUN, Norma is characterised by numerous substructures and exhibits an elongated X-ray morphology, implying that it is not yet in dynamical equilibrium (\citealp{Boehringer1996}, \citealp{Woudt2008}).
 
 ESO~137-006 (RA$_\mathrm{J2000}$ = 16:15:03.8, Dec$_\mathrm{J2000}$ = $-$60:54:26) is one of the most luminous galaxies in the Norma cluster and lies near the peak of the X-ray emission. It is one of the brightest radio galaxies in the southern sky (L$_{\rm 1.4 GHz} = 2.5\times 10^{25}$ W Hz$^{-1}$; \citealt{Sun2009}). Observations at 408 MHz \citep{schilizzi1975}, 843 MHz \citep{Mauch2003}, and 1400 MHz \citep{christiansen1977,Jones1996} show that it has a wide-angle tail morphology. The bending of its radio lobes is thought to be caused by the galaxy infall towards the main cluster \citep{Jones1996, Sakelliou2000}. In this letter we present new radio continuum images of ESO~137-006 at $\sim$1000 MHz and $\sim$1400 MHz based on MeerKAT observations, which reveal hitherto unseen collimated synchrotron threads (CSTs) between its radio lobes.\\

Throughout this paper we assume a $\Lambda$ cold dark matter cosmology with $\Omega_{\rm M} = 0.3$ and $\Lambda_{\Omega} = 0.7,$ and a Hubble constant of H$_{0}$ = 70 \kms\ Mpc$^{-1}$ . At the distance of ESO~137-006, 1\arcsec~corresponds to 0.33 kpc.

\section{MeerKAT observations and data reduction}\label{meerkatobs}
We observed the Norma cluster at radio frequencies with MeerKAT \citep{Jonas2016,mauch2019} in May 2019 (project ID SCI-20190418-SM-01). The observations were conducted with all 64 MeerKAT antennas in L-band (856 -- 1712 MHz) using the 4k mode of the SKARAB correlator, which samples the observed band with 4096 channels, each of 209 kHz in width, in full polarisation. The total integration time is 14 hours on target.

 \begin{figure*}
  \centering
 \includegraphics[width=150mm]{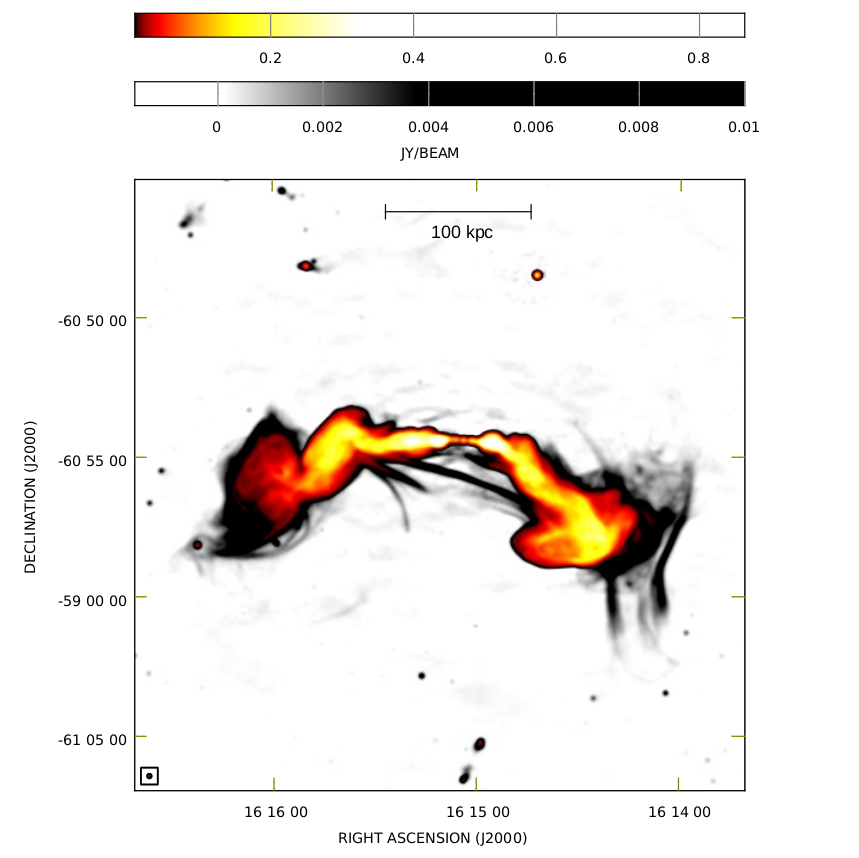}
 \caption{Radio continuum emission from ESO~137-006 detected by MeerKAT at 1030\,MHz. The top colour bar (yellow and red tones) represents the brightness of the brighter regions of the radio source in the range from 10 to 860\,mJy beam$^{-1}$. The bottom colour bar (grey tones) represents the brightness of the fainter plumes and filaments in the range from 10 down to -1.6\,mJy beam$^{-1}$.  The circular $\sim10$\arcsec\ synthesised beam of the image is shown in the bottom left corner. }\label{low_freq_hdr}
 \end{figure*}

 \begin{figure}
  \centering
 \includegraphics[width=90mm]{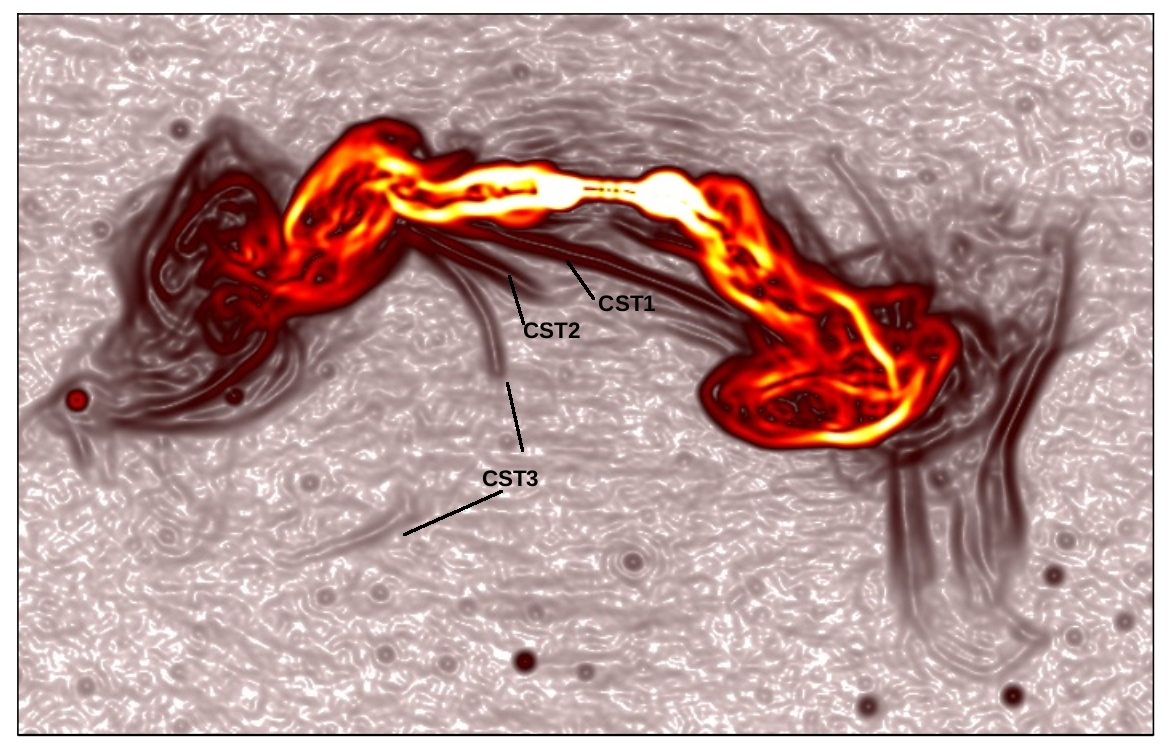}
 \caption{Gradient image (Sobel convolution kernel) with the three most prominent CSTs labelled.}\label{sobel}
 \end{figure}

We reduced the data in two frequency intervals largely free of radio-frequency interference: 980-1080 MHz and 1356-1440 MHz (hereafter referred to as 1030 MHz and 1398 MHz, respectively). The data reduction was conducted independently in the two intervals using the \textsc{caracal} pipeline currently under development\footnote{https://caracal.readthedocs.io}. The pipeline is built using \textsc{stimela}\footnote{https://github.com/SpheMakh/Stimela} \citep{makhathini2018}, a radio interferometry scripting framework based on Python and container technologies. \textsc{stimela} allows users to run several open-source radio interferometry software packages in the same script. 
Using this pipeline, we flagged the calibrator data based on the Stokes Q visibilities with \textsc{AOFlagger} \citep{Offringa2010}. We determined the complex flux, bandpass, and gains using the \textsc{casa} \citep{McMullin2007} tasks \textit{bandpass} and \textit{gaincal}, and applied the calibration to the visibilities of the target with the \textsc{casa} task \textit{mstransform}. The calibrated target visibilities were flagged with \textsc{AOFlagger}, again based on Stokes Q. We then iteratively imaged the radio continuum emission with \textsc{WSclean} \citep{Offringa2014} in Stokes\,I using multi-scale cleaning \citep{offringa2017}, and self-calibrated the gain phase with \textsc{cubical} \citep{Kenyon2018} with a solution interval of 128 seconds. The imaging was done using Briggs \it robust \rm value 0 and cleaning down to 0.5$\sigma$ within a clean mask made with \textsc{sofia} \citep{Serra2015}. Finally, we generated MeerKAT primary beam images at the mean frequency of the two processed bands using \textsc{eidos} \citep{Asad2019}, and created primary beam-corrected continuum images of the target.

The resulting 1030 MHz image has a restoring beam of $10.0\arcsec \times 9.1\arcsec$ FWHM with PA = $169\dg$, and rms noise level 30.8\,$\mu$Jy beam$^{-1}$. The 1398 MHz image has a restoring beam of $7.5\arcsec \times 6.8\arcsec$ FWHM with PA = $167\dg$, and rms noise level 20.8\,$\mu$Jy beam$^{-1}$.

\section{ESO~137-006 as seen by MeerKAT}
Figure\,\ref{low_freq_hdr} shows the 1030 MHz MeerKAT image of ESO~137-006. Emission shown in red to yellow is known from previous, shallower observations obtained with other telescopes (see Sec. \ref{sec:intro}). Emission shown in grey scale is revealed here for the first time due to the increased sensitivity and resolution of the MeerKAT data. A number of new features are now apparent.

\begin{figure}
  \centering
 \includegraphics[width=100mm]{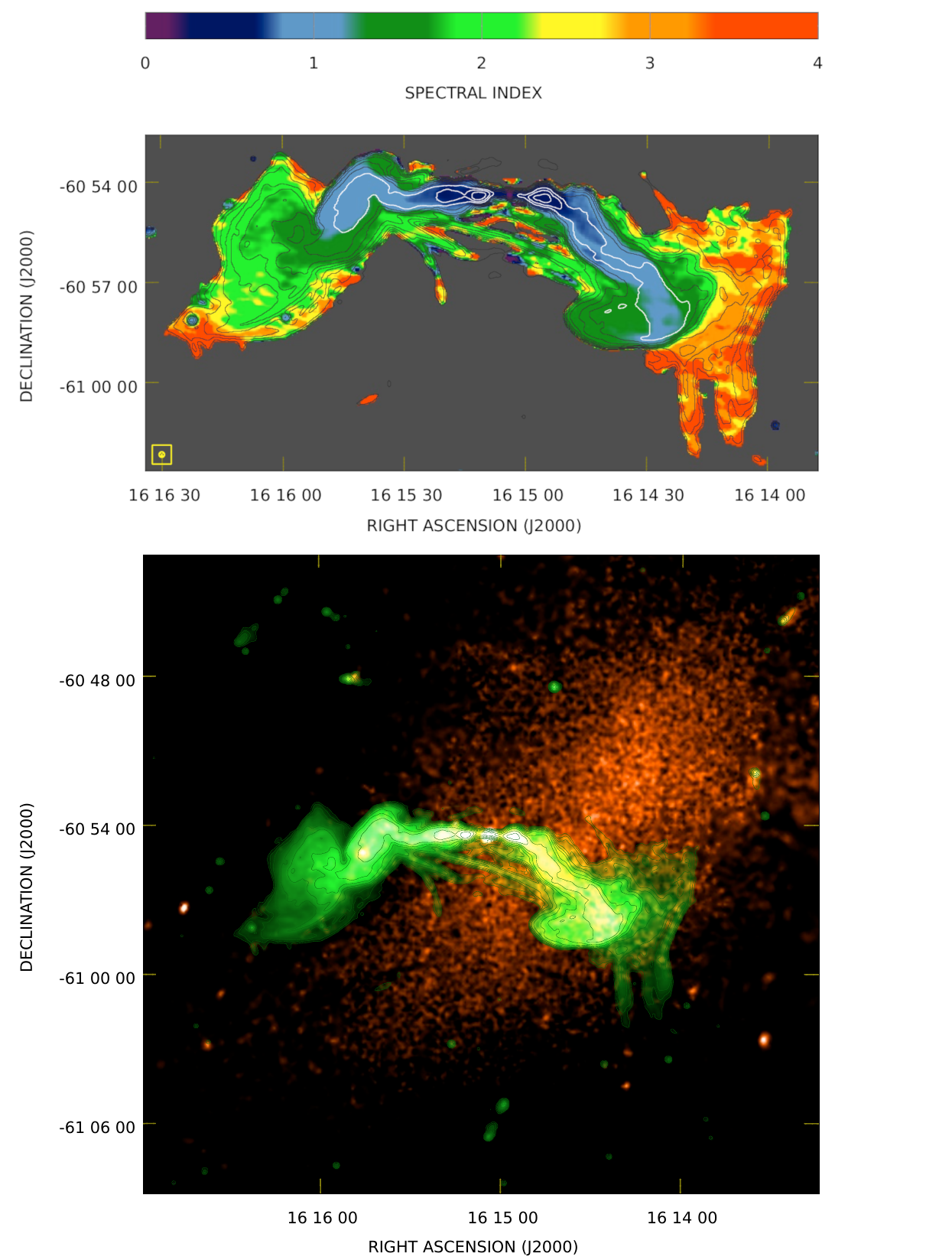}
 \caption{Top panel: Spectral index image between 1030 and 1398 MHz. Bottom panel: Radio continuum emission from ESO~137-006 detected by MeerKAT at 1030\,MHz (green tones and contours) superimposed on the X-ray emission map from XMM-Newton at 0.5 - 2 keV. In both panels, radio contours refer to the 1030 MHz image; they start at 0.5 mJy/beam and scale by a factor of two.}\label{xmm_and_meerkat}
  \end{figure}

Similar to earlier observations, we detect a point source corresponding to the galaxy ESO~137-006 \citep{Jones1996}. At the resolution of our images this source has a flux density of 140 mJy at 1030\,MHz and 167 mJy at 1398 MHz. The source is also seen in X-rays (see Fig.\,\ref{xmm_and_meerkat}) and is thought to be the core of ESO~137-006 (e.g., \citealt{Jones1996}). Narrow jets start from the core along an east--west axis and form two broad, bright spots $\sim1$\arcmin\ from it. Their peak flux densities are on average $\sim$8\,Jy beam$^{-1}$ at 1030 MHz and $\sim$18\,Jy beam$^{-1}$ at 1398 MHz. \citet{Jones1996} suggested that the brightening and expansion of the jets into these bright spots is due to the decreased ISM pressure. The general structure of these bright spots (see also Fig.\,\ref{sobel}) is similar to that seen in the hydrodynamical simulations of jets in the FR\,I radio galaxy 3C\,31, where the jets recollimated at $\sim$1.5 kpc from the galaxy core due to the decreased pressure relative to the ISM after their initial expansion \citep{Perucho2007}.
 
Further out, the jets flare into the lobes and then fade into diffuse emission. On both sides, the radio emission bends southwards; in the western lobe the bending occurs right after the bright spot while in the eastern lobe it occurs further out.

Besides the previously known features described immediately above, Fig.\,\ref{low_freq_hdr} shows additional, low-surface-brightness radio emission at the edge of both lobes and, most strikingly, a number of CSTs in the region south of the core. These extend from the lobes southward of the core. Figure\,\ref{sobel} shows the gradient image obtained with a Sobel convolution kernel that highlights the presence of three threads: CST1, CST2, and CST3. These threads are reminiscent of twin jets originating from two nuclei associated with radio source 3C\,75 (\citealp{Owen1985}), except for the fact that in the case of ESO~137-006 only one nucleus is seen.
Figure\,\ref{low_freq_hdr} also shows a diffuse loop of emission which starts at the base of CST3, extends around the eastern lobe on the south side, and connects back to the lobe on its easternmost side where it fades into diffuse emission. The origin of these features is not known. It is possible that these loops and CSTs are the result of some electromagnetic effect like parallel magnetic field lines within the ICM dragging particles between the lobes as the galaxy moves north, perhaps similar to the description by \citet{Heyvaerts1989}.

The spectral index image ($S_{\nu}\propto \nu^{-\alpha}$) of ESO~137-006 between 1030 and 1398\,MHz is presented in the top panel of Fig.\,\ref{xmm_and_meerkat}. We find a spectral index distribution typical of an active wide-angle tail radio source. The radio core has a flat spectral index $\alpha\simeq 0$ while the inner and brighter parts of the jets have $\alpha\simeq 0.5 - 0.6$. There is a steepening in the radio spectrum going from the radio lobes down to the tails where we measure a spectral index as high as $\alpha\simeq 4$. The spatial variation of the spectral index in the eastern lobe is more gradual than in the western lobe, where a sharp transition, likely due to projection effects, is observed between the edge of the lobe and the underlying tail emission. The newly discovered CSTs are characterised by a steep radio spectrum with $\alpha\simeq 2$. 

In the bottom panel of Fig. \ref{xmm_and_meerkat} we present the MeerKAT image at 1030 MHz superimposed on the X-ray emission map from XMM-Newton at 0.5 - 2 keV. The most relevant feature seen in the X-ray image is a point-like source associated with the radio core and a hint of cavities corresponding to the radio lobes, although no evidence for cavities associated with this source is present in the literature \citep{Shin2016}. At the sensitivity limit of the XMM image, there are no X-ray features that we can associate with these CSTs.

\begin{figure}
  \centering
 \includegraphics[width=90mm]{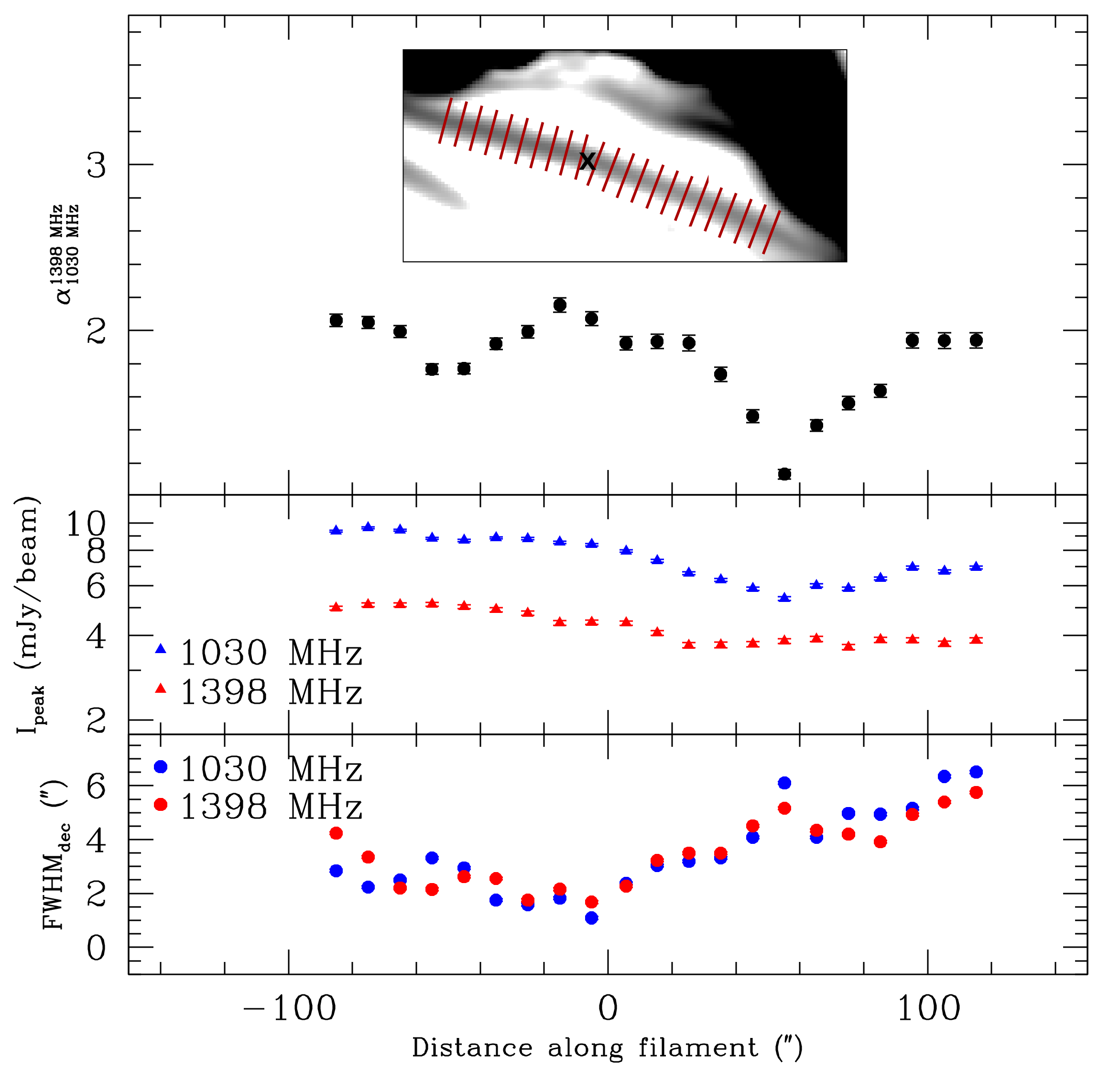}
 \caption{Top panel: Spectral index profile between 1030 and 1398\,MHz computed along the filament at the locations shown by the slices in the inset (see text for details). Middle panel: Peak brightness at 1030\,MHz (blue triangles) and at 1398\,MHz (red triangles) along the filament at the location of the slices. Bottom panel: Deconvolved FWHM at 1030\,MHz (blue dots) and at 1398\,MHz (red dots) along the filament at the location of the slices. The plotted error bars are comparable to the size of the dots.}\label{profile}
 \end{figure}

\section{Basic properties of CST1}
Figure\,\ref{profile} shows the deconvolved full width half maximum ($\rm FWHM_{\rm dec}$) and the peak brightness ($\rm I_{\rm peak}$) as a function of position along CST1 at 1030 and 1398\,MHz, as well as the spectral index ($\alpha_{\rm 1030\,MHz}^{\rm 1398\,MHz}$). We measure these quantities at the positions shown by the red bars in the inset. The cross in the inset corresponds to the origin of the plot's horizontal axis. Errors were computed using a Monte Carlo simulation with input parameters sampled and noised in the same way as for the real data. The errors represent the scatter of the input versus output best-fit parameters. The mean values of the peak brightness are respectively $\rm I_{\rm peak,1030\,MHz}= (8 \pm 1)\,{\rm mJy/beam}$ and $\rm I_{\rm peak,1398\,MHz}= (4.3 \pm 0.6)\,{\rm mJy/beam}$, and the average spectral index $\alpha_{\rm 1030\,MHz}^{\rm 1398\,MHz}=1.8\pm0.3$, consistent with the values shown in Fig.\,\ref{xmm_and_meerkat}.

The peak brightness shows a similar profile at both frequencies, with a valley at $\sim 60^{\prime\prime}$ at 1030\,MHz only, which corresponds to a depression in the spectral index profile. At the same location, a rapid local increase of the deconvolved FWHM can be observed at both frequencies. The mean width of CST1 is $\rm FWHM_{\rm dec, 1030\,MHz}=(3.5\pm1.6)^{\prime\prime}$ and 
$\rm FWHM_{\rm dec, 1398\,MHz}=(3.5\pm1.2)^{\prime\prime}$. The two profiles do not show any dependence on frequency but are rather characterised by a similar average trend along the full extension of the filament, consistent within the scatter. This projected width translates into $1.2\pm 0.5$ kpc, or $\sim 2$\% of its 80\,kpc length.

\section{Discussion and summary}

In this letter we present new MeerKAT images of the radio source ESO~137-006 at 1000 and 1400 MHz. The galaxy lies at the centre of the merging Norma cluster near the Great Attractor. Here we summarise our main findings:

\begin{itemize}
    \item With these sensitive MeerKAT observations, new  features have been revealed in the form of multiple collimated synchrotron threads (CSTs) connecting the lobes of the radio galaxy. 
    It is worth noting that examples of filamentary structures associated with radio galaxies are well known in the literature. However, these filaments are usually observed \emph{inside} the radio lobes (see e.g. the notable cases of Fornax A and Cygnus A; \citealp{Maccagni2020}, \citealp{Perley1984}) and the tails of radio galaxies (e.g.,  NGC~1265, 3C~129, and NGC~326; \citealp{Sijbring1998}, \citealp{Lane2002}, \citealp{Hardcastle2019}). The CSTs detected in ESO~137-006 are different in that they are observed \emph{outside} the main body of the radio galaxy and connecting (at least in projection) the two radio lobes. The radio galaxy 3C~338 (\citealp{Burns1983}) at the centre of Abell 2199 presents a single filament that is reminiscent of one of the CSTs observed in ESO~137-006. However, while the filament in 3C~338 could be a relic jet from a past epoch of activity, this same interpretation does not hold for ESO~137-006 where we observe \emph{multiple} close-by threads formed at the same time (as suggested by their similar spectral-index distributions).
    
    \item The most prominent and straight of the CST in ESO~137-006 (CST~1) has a characteristic width of $\sim 1$\,kpc (deconvolved FWHM), roughly 2\% of its length, and has a relatively smooth brightness profile with a peak intensity of a few mJy/beam at the 10\arcsec \ resolution of our images. The other two CSTs originate from the same point in the eastern lobe. CST~2 starts straight and then fades rapidly after $\sim25$ kpc from the lobe. CST~3 seems to follow a faint, closed loop with a radius of  $\sim 64$ kpc, which reconnects with the lobe at its far end. The nature of these unusual features is unclear. We speculate that they could be due to the interaction of the magnetic fields of the  radio lobes  with the magneto-ionic ICM, or caused by some sort of re-connection of filaments associated with the tails back into the radio lobes. Further observations and theoretical efforts are required to clarify the nature of these newly discovered features.
    
    \item The spectral index distribution observed across the jets, lobes, and tails is typical of an active radio source. The radio spectrum of the CST is steep with $\alpha\simeq 2$. Due to this steep spectrum, deep low-frequency observations at high resolution with instruments such as the LOw Frequency ARray could play a role in the study of CSTs. 
 
    \item Whatever their origin, our findings pose the following questions: How common are these features? Are CSTs specific to the case of ESO~137-006 and its environment, perhaps due to the kinematics and pressure gradient of this ICM and relative motion of the galaxy in the cluster? Or, on the contrary, are CSTs common in radio galaxies but have so far not been detected due to sensitivity and resolution limits?
    If future observations confirm the latter hypothesis, understanding the nature and the physics of these features could open a new science case for the next generation of sensitive radio interferometers like the Square Kilometre Array. 
    
\end{itemize}

\begin{acknowledgements}
We thank the referee, Manel Perucho for the useful comments and suggestions. The authors also wish to thank Ming Sun for helping with the access to some of the X-ray data. This project has received funding from the European Research Council (ERC) under the European Union's Horizon 2020 research and innovation programme grant agreement no. 679627 project name FORNAX. MR acknowledges support from the Italian Ministry of Foreign Affairs and International Cooperation (MAECI Grant Number ZA18GR02) and the South African Department of Science and Technology's National Research Foundation (DST-NRF Grant Number 113121) as part of the ISARP RADIOSKY2020 Joint Research Scheme. This paper makes use of the MeerKAT data (Project ID: SCI-20190418-SM-01). The MeerKAT telescope is operated by the South African Radio Astronomy Observatory, which is a facility of the National Research Foundation, an agency of the Department of Science and Innovation. MR and SM's research is supported by the SARAO
HCD programme via the "New Scientific Frontiers with Precision Radio Interferometry" research group grant. This work is based upon research supported by the South African Research Chairs Initiative of the Department of Science and Technology and National Research Foundation.
(Part of) the data published here have been reduced using the CARAcal pipeline, partially supported by BMBF project 05A17PC2 for D-MeerKAT. Information about CARAcal can be obtained online under the \url{https://caracal.readthedocs.io/en/latest/}.
\end{acknowledgements}

\normalsize
\bibliographystyle{aa} % 
\bibliography{AA_2020_37800_Ramatsoku.bib} % your file with extension .bib containing references

\end{document}